\documentclass[12pt]{iopart}
\usepackage{graphicx}
\usepackage{hyperref}
\usepackage{color}

\begin{document}

\title{Probing into the effectiveness of self-isolation policies in epidemic control}

\author{Nuno Crokidakis $^{1,2}$ and S\'{\i}lvio M. Duarte Queir\'os$^{3}$}
\address{
$^{1}$ Departamento de F\'isica, PUC-Rio and National Institute of Science and Technology for Complex Systems \\
Rua Marqu\^es de S\~ao Vicente 225 \\ 22451-900 Rio de Janeiro - RJ, Brazil \\
$^{2}$ Instituto de F\'{\i}sica, Universidade Federal Fluminense \\
Av. Litor\^anea s/n \\
24210-340 Niter\'oi - RJ, Brazil \hspace{15mm} \\
$^{3}$ Istituto dei Sistemi Complessi, CNR \\
Via dei Taurini 19 \\ 00185 Rome, Italy}

\ead{nuno.crokidakis@fis.puc-rio.br, sdqueiro@gmail.com }

\begin{abstract}

In this work, we inspect the reliability of controlling and quelling an epidemic disease mimicked by a Susceptible-Infected-Susceptible (SIS) model defined on a complex network by means of current and implementable quarantine and isolation policies. Specifically, we consider that each individual in the network is originally linked to two types of individuals: members of the same household and acquaintances. The topology of this network evolves taking into account a probability $q$ that aims at representing the quarantine or isolation process in which the connection with acquaintances is disrupted according to standard policies of control of epidemics. Within current policies of self-isolation and standard infection rates, our results show that the propagation is either only controllable for hypothetical rates of compliance or uncontrollable at all.
\end{abstract}

%\pacs{05.10.Ln, 05.50.+q, 64.60.De, 75.10Hk, 75.40.Mg}

\maketitle

\section{Introduction}

\qquad Alongside with wars and natural disasters, plagues and epidemic (pandemic) diseases lie at the top of death toll lists in Human history. By directly interfering with the risk of death, such events or even their likelihood are the source of acute distress among populations, which has been well-documented by historians since ancient times~\cite{procopius}. Concerning plagues and epidemic diseases, isolation of infected people has been applied, at least, since Old Testament period as an instrument for controlling and quelling the spread of both viruses and contamination agents in such situations~\cite{bible}. If for centuries it was possible to estrange people from their relatives and whereabouts by means of a simple decree, social conquests based on the ``Declaration of the rights of man and of the citizen'', which led to ``Universal declaration of Human rights'', have urged the discussion over the morality and legitimacy of compulsory quarantine and isolation (Q\&I) practices~\cite{swendiman,jones_shimabukuro, swendiman_jones,coleman, who}. Indeed, countries like Brazil and Japan have repealed previous laws for compulsory isolation, abortion and sterilisation of patients suffering from leprosy as well as decided to pay compensations and allowances to people subject to these practices~\cite{brooke,brazil,zulino}.  Accordingly, current Q\&I practices strongly rely on the consciousness of the infected individual and her attitude towards the rest of the society by imposing a self-isolation spell according to a medical recommendation (see~\cite{who} and references therein). Concomitantly, recent polls have shown that the willingness to comply with a self-isolation period depends on the social condition and literary habilitation of the individual very much~\cite{eastwood}.

Quantitatively, the problem of reasoning over the spreading of an infectious and epidemic disease is generally based on standard models such as the Susceptible-Infected (SI), the Susceptible-Infected-Susceptible (SIS) and  the Susceptible-Infected-Recovered (SIR) models~\cite{anderson_may,bailey} and their variants. Despite the simplicity of these models, they have been successfully applied to a variety of cases~\cite{diekmann,esteva_vargas,haye,schwartz,baryarama,gordo, katriel,nuno} in the assessment of the propagation of an epidemic disease. In the form of differential equations, their dynamics is  well-known in the literature of the theme where they are usually called general solutions. When geometrical conditions are taken into account, namely the structure of the network of relationships between people, the critical behaviour changes~\cite{boguna_2003,castellano}. Within this context, analytical and numerical results concerning the existence or not of a non-zero transition for quenched scale-free networks has been in the spotlight~\cite{chatterjee_durrett,ganesh,krivelevich,may_lloyd,parshani,pastor-satorras1,pastor-satorras2}.

In our case, we have opted to base our model on the SIS model, which can represent situations wherein the virus responsible for the infection is able to mutate fast. Explicitly, although someone is recovered from the infection, she is at risk of being infected once again due to the newest mutation. Concerning the topology in which the phenomenon endures, we have assumed a dynamical complex network with $N$ nodes, each one linked to $k$ other nodes in accordance with a distribution $P(k)$. Each node of the network represents an individual either in a Susceptible, $S$, or in an Infected, $I$, state. As usual, we start the dynamics with a single infected individual, randomly chosen among the $N$ nodes. At first, all the remaining individuals are Susceptible. In addition, we have considered that a density of the connections,  $d$, in the network are fixed and thus they cannot be removed during the Q\&I process. Two people connected by a fixed link are defined as members of the same household, whereas people linked by non-fixed links are dubbed acquaintances.

Recently, some works have studied the effect of isolation in the physics and computer science literature~\cite{gross,lagorio,shaw1,shaw2,wang,mukamel,marceal}. It was therein argued that such policies can be effective in controlling the spread of an epidemic disease. However, the conditions of these models rely on the total isolation of the infected patient, which is closer to academic proposals that also inspired several literary contemporary masterpieces~\cite{camus,saramago} rather than an actual implementable policy following international rules or recommendations. Alternatively, total isolation scenarios have also been studied in the context of small islands. In this case, it is argued that these policies might be successful, but in quite stern conditions demanding a complex and expensive logistics~\cite{garcia,nishiura}. On the other hand, the study of the impact of the household in the propagation of SIR epidemics as well as adequate vaccination policies have been discussed in a biological and medical framework~\cite{ball_neal_2002,ball_lyne,ball_neal_2008,becker_dietz,pellis,fraser}. However, for the sake of analytical treatment~\cite{pellis}, they often neglect the structure of social networks and assume a random contact approach between individuals. Complementary, surveys stemming from the analysis of data on fatalities due to the ``Spanish flu'' pandemic in cities of the United States of America and its relation to public health and non-pharmaceutical interventions have been presented~\cite{hatchett,ferguson_bootsma}. In spite of the fact that both results provide important insights into the relevancy of coordinated interventions and respective set about, (American) society has dramatically changed in almost $100$ years and thus its epidemic response to such interventions. Moreover, some of the measures adopted during that pandemic are now liable of being judged unethical or breaching some fundamental law, as we have previously mentioned to. Last but not least, it was impossible in these studies to separate out the impact of each measure.

Thence, with this work we intend providing an answer to the fundamental question about to what extend Q\&I policies abiding by World Health Organization (WHO) directives, which are prone to self-conscious isolation, are real hinderers of the spreading of an epidemic disease, contributing  in this way to its control and suppression. Quantitatively the task of answering to this question is presented in the form of looking for an infection rate (or epidemic) threshold, $\lambda^{\ast}$, for a given rate of effectiveness of the isolation, above which the disease persists or it is quelled otherwise.

% ###################################################################

\section{Model and Numerical Results}

\qquad At each time step, $t$, the following automata rules control our model:
\begin{itemize}
\item We visit every node in the network;

\item For each Infected node, we look for all Susceptible neighbours of hers;

\item For each $I-S$ link connecting members of the same household, the Susceptible individual becomes Infected with probability $\lambda$;

\item On the other hand, we remove each $I-S$ link connecting acquaintance individuals with probability $q$ (quarantine). Then, the Infected individual is isolated and the Susceptible one is reconnected to another randomly chosen Susceptible node, who is not a current contact of hers. For each surviving $I-S$ link, the Susceptible individual becomes Infected with probability $\lambda$;

\item After the verification of all neighbours of an Infected individual, she returns to the Susceptible state with probability $\alpha$.

\end{itemize}

%%%%%%%%%%%%%%%%%%%%%%%%%%%%%%%%%%%%%%%%%%%%%%%%%%%%%%%%%%%%%%%%%%%%%%%%%%
\begin{figure}[t]
\begin{center}
\vspace{1.0cm}
\includegraphics[width=0.45\textwidth,angle=0]{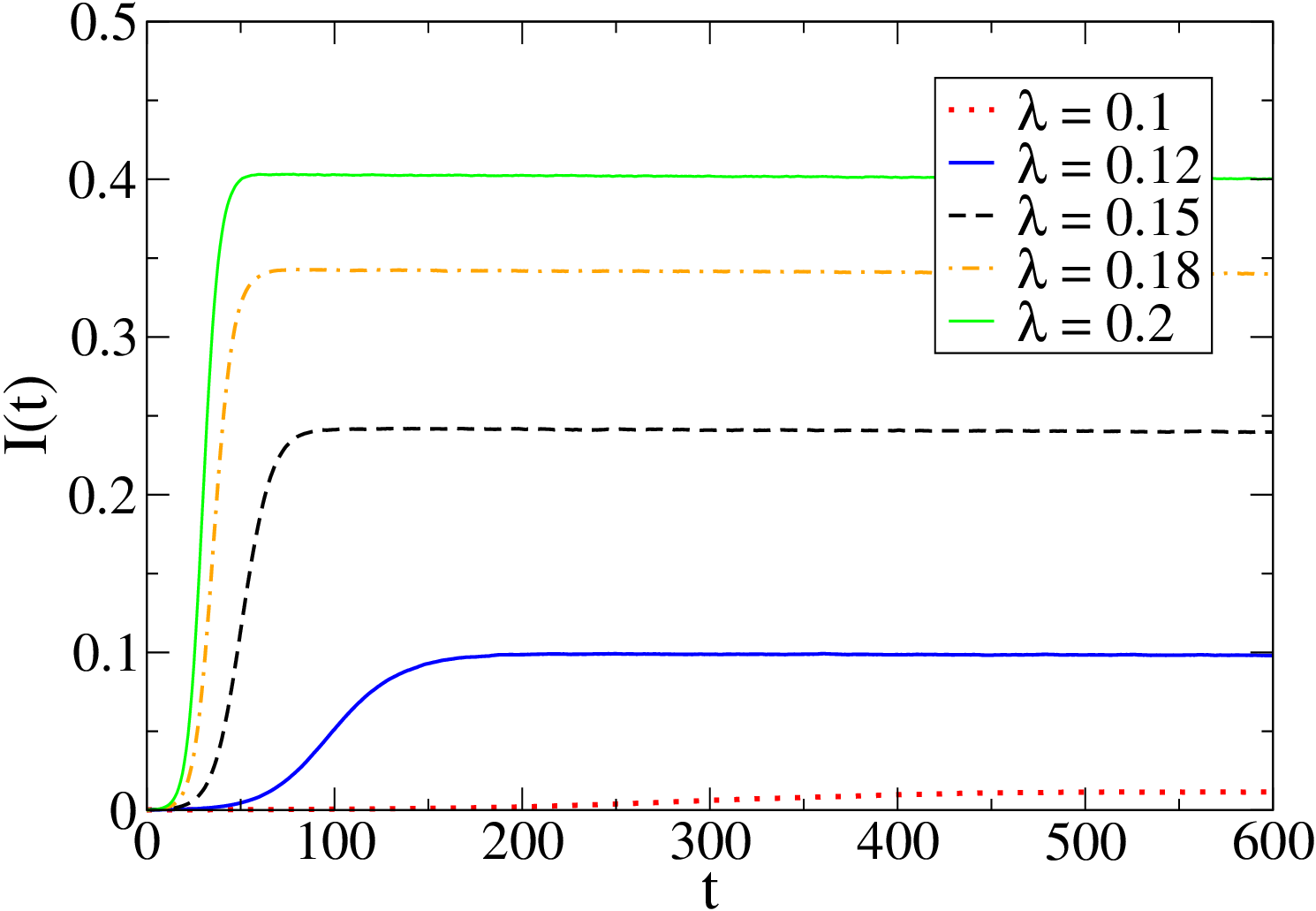}
\hspace{0.5cm}
\includegraphics[width=0.45\textwidth,angle=0]{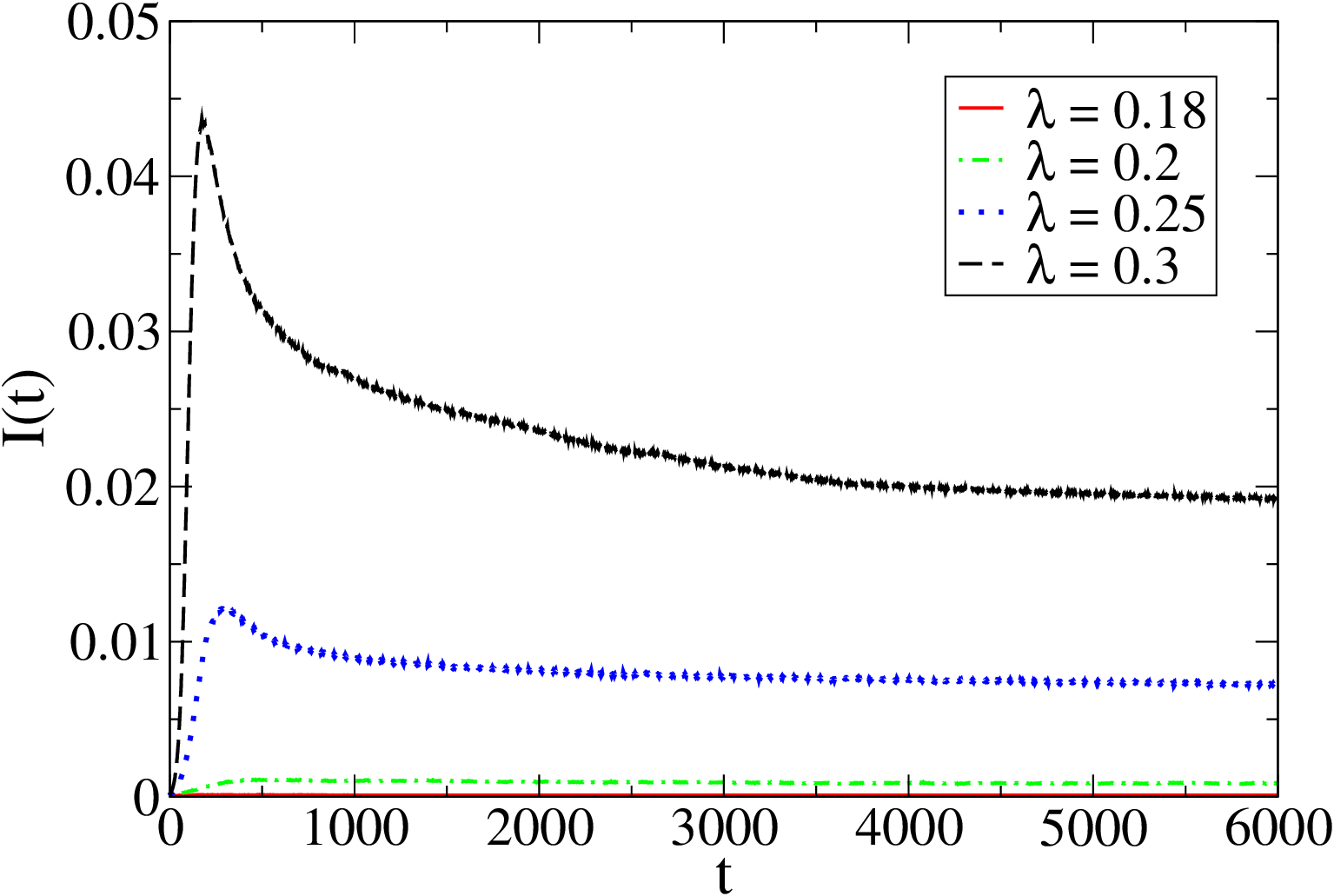}
\end{center}
\caption{Time evolution of the density of Infected individuals for $d=0.2$ and $q=0.3$ (left side) and $q=0.8$ (right side), and typical values of the infection probability $\lambda$ for random networks with $N=10^{4}$ nodes and $\langle k\rangle=5$. Notice that the time needed for the system to reach the steady states increases for increasing values of the self-isolation probability $q$  of the Q\&I process. Results are averaged over $10^{3}$ realisations.}
\label{fig1}
\end{figure}
%%%%%%%%%%%%%%%%%%%%%%%%%%%%%%%%%%%%%%%%%%%%%%%%%%%%%%%%%%%%%%%%%%%%%%%%%%%

Let us elaborate upon the parameters and the rules we have established. We start by discussing the role of $q$ the main purpose of which is to reproduce the probability that a patient complies with the medical recommendation to stay home. Ideally, people would strictly abide by the physician's counsel, but as recent polls have showed and given statistical significance, the rates of compliance are different from $100 \% $ and can go as low as $50 \% $. Two main factors for this change have been identified~\cite{eastwood,kavanagh}; {\it i)} More often than not people do not understand or distort what they are expected to do due to literacy skills~\cite{kavanagh}; {\it ii)} As occurs in life in general, people (patients included) perform a risk assessment~\cite{eastwood} before making a decision on stopping contact with a given individual. Accordingly, infected agents decide to disrupt the contacts they have depending on the time they were asked to stay home, the type, namely the importance (either circumstantial or not) of the relationship between the infected person and the susceptible individual, and the number of people that can be involved when the attendance of a given event is considered. Furthermore, this risk assessment is weighed by the hazard that the eventual non-compliance carries. Effectively, the perception of risk is quite affected as the time of the self-isolation goes by as well as the reckoned performance of pharmaceutical interventions. As a matter of fact, glaring differences in the polls results concerning self-isolation before and after the introduction of a vaccine in the H1N1 pandemics were verified. A similar behaviour has been verified in polls on high risk unprotected sexual relations as well~\cite{risk}.

Moving ahead, we now underpin the introduction of the rewiring process. As more than well known, social relationships come to pass because people need one another. Besides friendship, familiar and sentimental ties, people relate with other people because of their functions, skills and responsibilities. Previous studies have considered that after a link between susceptible and infected individuals having been destroyed, no new relation is established. In a sociological perspective, this option acts as though the needs of the susceptible individual can be solved within her current circle. Our option goes along another path we consider more realistic and that is close to processes of rewiring studied in other works~\cite{gross,gross_blasius_2008,prx}. In spite of the fact that we could have considered a probability $f$ that a susceptible ``stranded'' agent relinks, we have opted to mitigate the number of parameters of the problem and keep it constant and equal to 1.\footnote{We can assume that neither her householders nor acquaintances manage to act as a stand-in.}

As the majority of people live together, in very few (and negligible) cases, it is possible to someone to really isolate herself from the people with whom she shares an address. In this way, we justify the introduction of parameter $d$. Still, we would like to shed light on some aspects of this parameter. First, its introduction does not transform our network into a superposition of two networks (householders and acquaintances).\footnote{From a coarse grained point of view, we could use this approach to define a network households but this is not a superposition of networks as well since in this case each node is now a household and not an individual.} The network remains unique, but there is a fraction $d$ of the links that cannot be changed because they have been quenched. Householders are the people who share the same residence whether they are relatives or not (e.g., student's flats) and acquaintances are people with whom someone maintains contact with in a close and regular basis representing at its best a fraction of the so called Dunbar's number, which is an anthropological/cognitive measure that quantifies the number of people with whom each person is able to keep on a relationship, \emph{i.e.}, preserving some information about his life. It is also important emphasise that in establishing the network we have borne in mind important details, \emph{e.g.}, if two people are both members of the same household of a third individual, then a quenched connection between them is immediately established because they are obviously members of the same household.

As regards other computational details, we have applied a synchronous update, \emph{i.e.}, after having visited all the nodes on the network, we update all the individuals' states simultaneously~\cite{grassberger,hinrichsen}. In other words, a new time step is considered only when the state of all the individuals has been scrutinised. Furthermore, we have considered in this work a fixed $I\to S$ transition probability $\alpha=0.2$ and we have simulated the model on a Erd\"{o}s-R\'enyi network with $N=10^{4}$ nodes and $\langle k \rangle = 5$. The results herein presented have been obtained averaging over $10^3$ independent simulations, for each set values presented. Please observe we have only considered the epidemic spreading in the largest connected component of the network, \emph{i.e.}, the giant cluster \cite{lagorio}.

%%%%%%%%%%%%%%%%%%%%%%%%%%%%%%%%%%%%%%%%%%%%%%%%%%%%%%%%%%%%%%%%%%%%%%%%%%
\begin{figure}[t]
\begin{center}
\vspace{1.0cm}
\includegraphics[width=0.45\textwidth,angle=0]{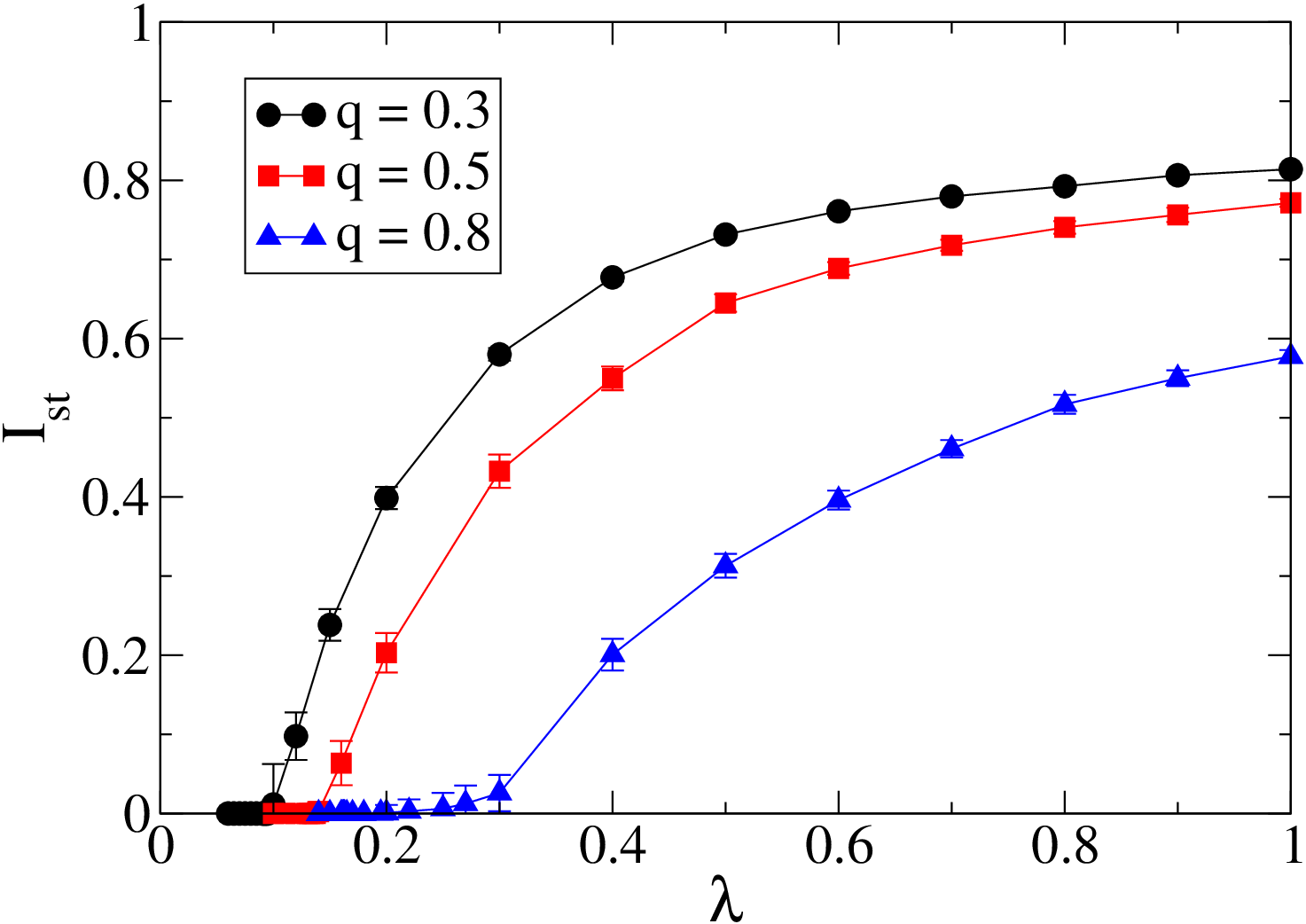}
\hspace{0.5cm}
\includegraphics[width=0.45\textwidth,angle=0]{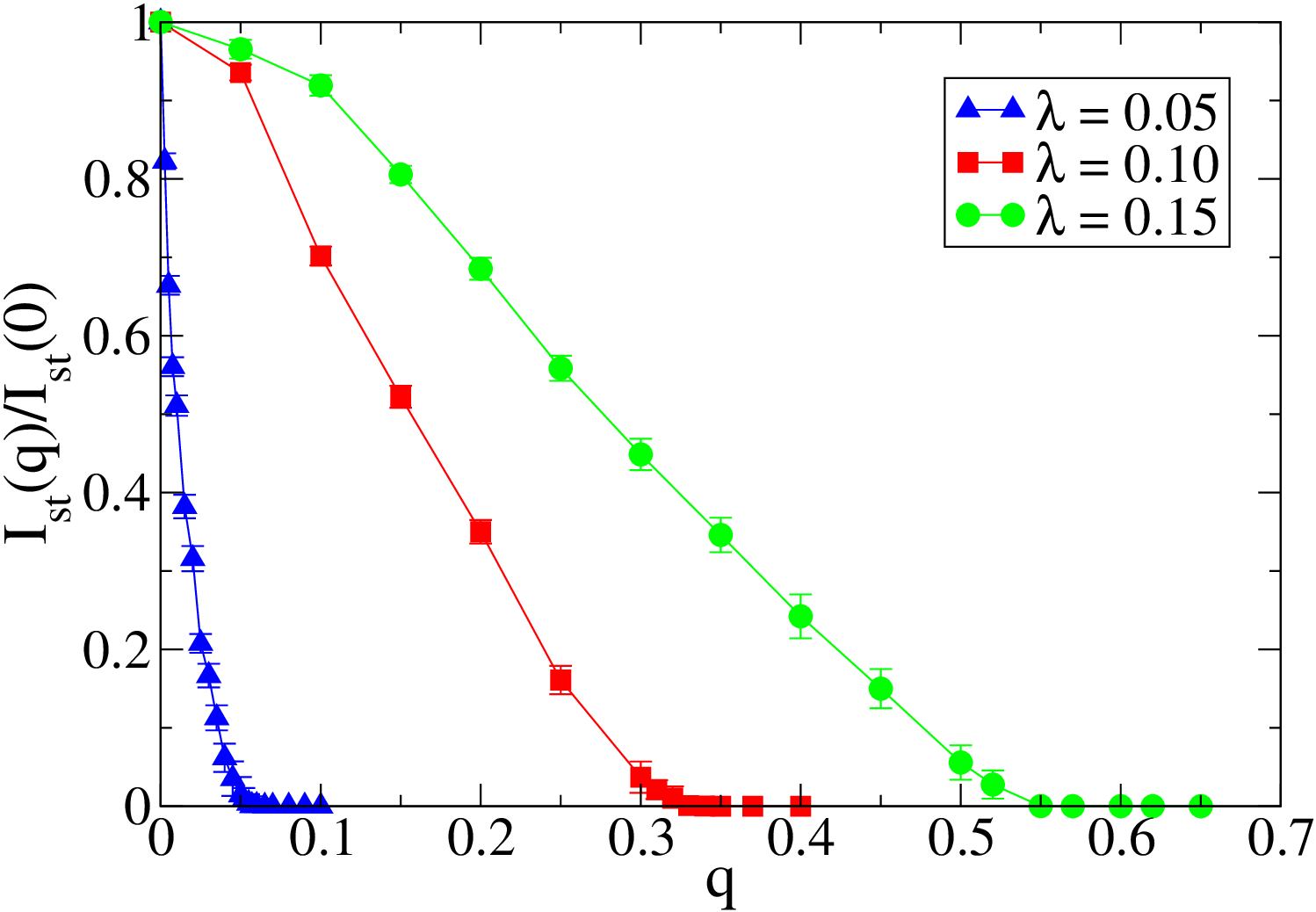}
\end{center}
\caption{\textit{Left side:} Normalised stationary density of Infected individuals $I_{st}$ as a function of $\lambda$ for $d=0.2$ and typical values of $q$. We can observe a transition from $I_{st}=0$ to $I_{st}>0$ at different threshold points $\lambda(q)$. \textit{Right side:} $I_{st}$ as a function of the self-isolation probability $q$ of the Q\&I process, for $d=0.2$ and typical values of the infection probability $\lambda$. We also observe a transition for each value of $\lambda$, at different points. In all simulations, $N=10^{4}$, $\langle k \rangle=5$ and averages are over $10^{3}$ realisations.}
\label{fig2}
\end{figure}
%%%%%%%%%%%%%%%%%%%%%%%%%%%%%%%%%%%%%%%%%%%%%%%%%%%%%%%%%%%%%%%%%%%%%%%%%%%

In Fig.~\ref{fig1}, we exhibit the time evolution of the density of Infected individuals $I(t)$ for $20\%$ of fixed links in the network (i.e., for $d=0.2$), two different values of the self-isolation probability, namely $q=0.3$ and $q=0.8$, and $\lambda $ scanning the domain of infection rates. As it can be observed, there are different values of $\lambda$ above which we have $I_{st}>0$. Here, the notation $I_{st} \equiv I(t\to\infty)$ stands for the stationary density of Infected individuals.\footnote{The quantity $I_{st}$ was obtained from time averages of $I(t)$ taken in the steady state, and in addition we also considered averages over $10^{3}$ network realisations.} This defines the usual phase transition of epidemic models: for $\lambda\leq \lambda^{\ast}(q)$, we have a disease-free phase with all the individuals presenting a $S$ state, whereas for $\lambda> \lambda^{\ast}(q)$ we have an epidemic phase, \emph{i.e.}, the disease spreads out and a finite fraction of the population is constantly infected. Pay heed to the fact that, for increasing values of $q$, the time needed for the system to reach a stationary state increases. This kind of transition is best observed in Fig.~\ref{fig2} (left panel), wherein we exhibit the stationary density of Infected individuals $I_{st}$ as a function of $\lambda$ for $d=0.2$ and typical values of the self-isolation probability $q$ of the Q\&I process. In this case, we can see that for a fixed value of $q$ the transition occurs at different values of $\lambda$, which defines the transition points $\lambda^{\ast}$. Analogously, the stationary density of Infected individuals $I_{st}$ can be represented as a function of the self-isolation probability $q$ of the Q\&I process for different values of $\lambda$ (see Fig. \ref{fig2}, right panel). Once more, we perceive a transition at different values of $q(\lambda)$. Notice that when we increase the self-isolation probability, the disease only disseminates through the network if we increase the infection probability $\lambda$. In other words, the final size of the epidemic may be reduced, as was observed in the 2009 H1N1 pandemic \cite{eastwood} and as also discussed in some works \cite{gross,lagorio,shaw1,shaw2,gross_blasius_2008}. Nonetheless, in what follows, we will conclude its outcome is limited when we take into account a more realistic model with a density of fixed links.

%%%%%%%%%%%%%%%%%%%%%%%%%%%%%%%%%%%%%%%%%%%%%%%%%%%%%%%%%%%%%%%%%%%%%%%%%%
\begin{figure}[t]
\begin{center}
\vspace{1.0cm}
\includegraphics[width=0.5\textwidth,angle=0]{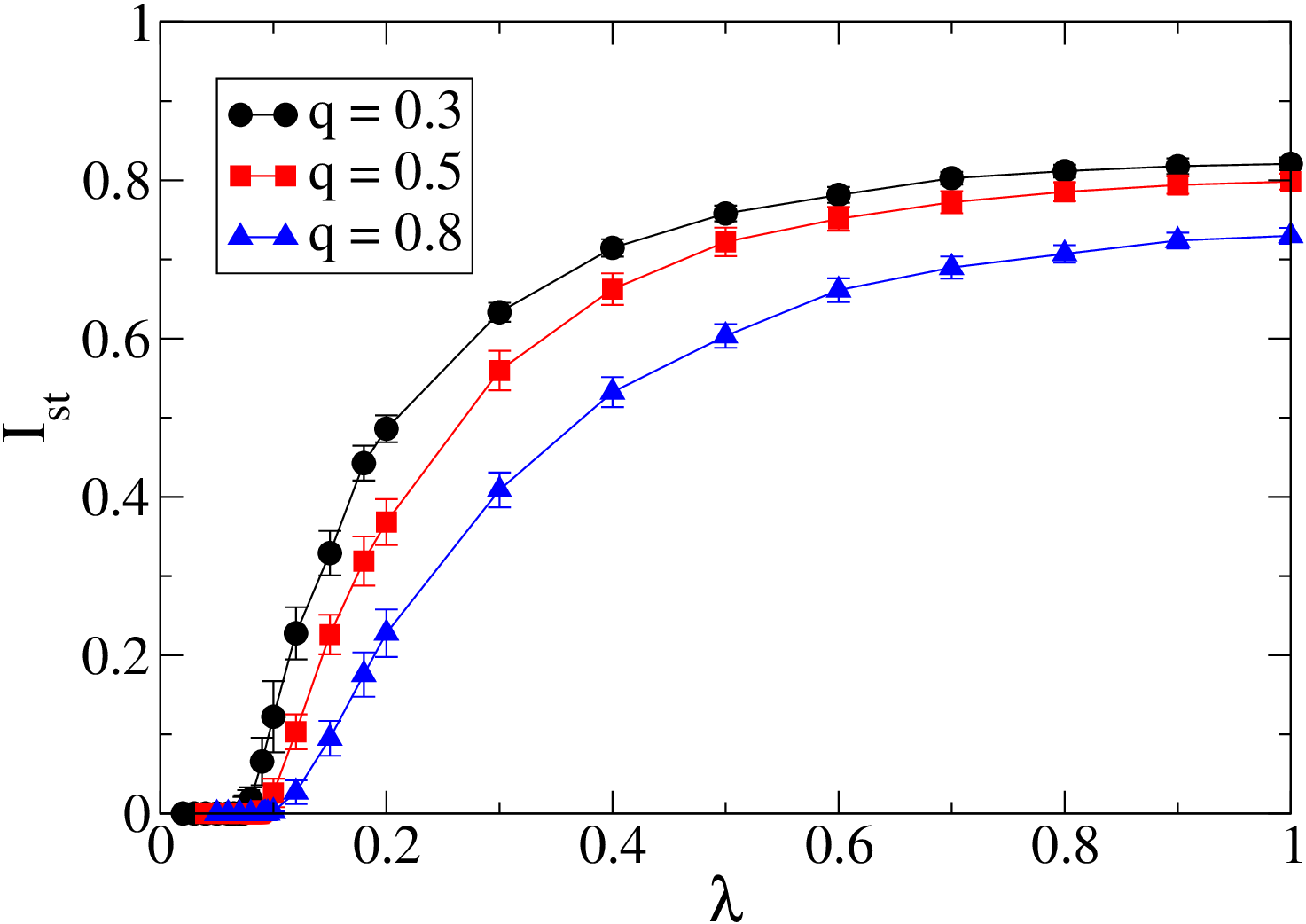}
\end{center}
\caption{Normalized stationary density of Infected individuals $I_{st}$ as a function of $\lambda$ for $d=0.4$ and typical values of $q$. We can also observe a phase transition from $I_{st}=0$ to $I_{st}>0$ at different threshold points $\lambda(q)$. In all simulations, $N=10^{4}$, $\langle k \rangle=5$ and averages are over $10^{3}$ realizations.}
\label{fig3}
\end{figure}
%%%%%%%%%%%%%%%%%%%%%%%%%%%%%%%%%%%%%%%%%%%%%%%%%%%%%%%%%%%%%%%%%%%%%%%%%%%

\begin{figure}[t]
\begin{center}
\vspace{1.0cm}
\includegraphics[width=0.5\textwidth,angle=0]{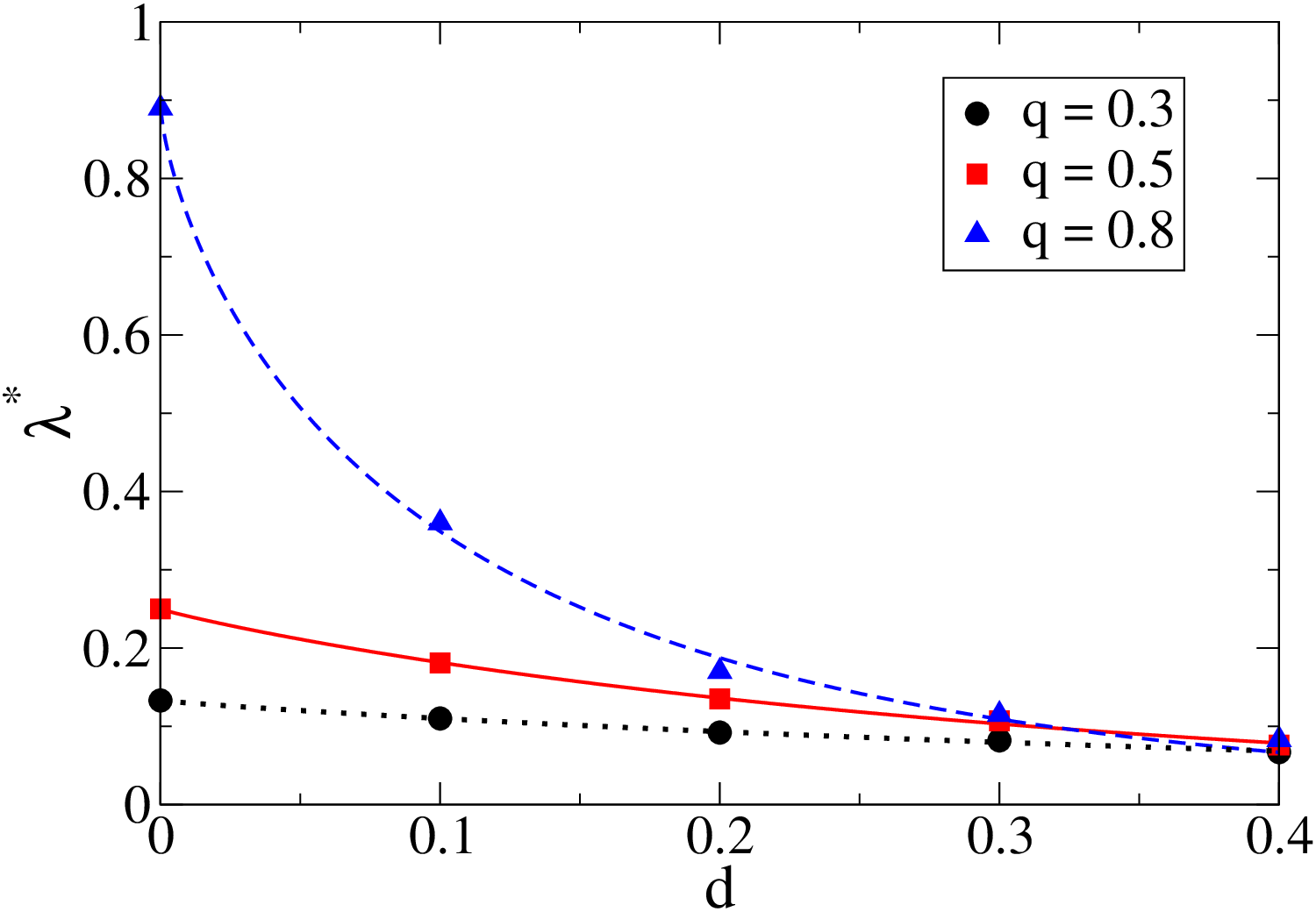}
\end{center}
\caption{The infection probability threshold $\lambda^{\ast}$ as function of $d$ for typical values of the self-isolation probability $q$ of the Q\&I process. The lines are fittings based on the stretched exponential function $\lambda^{\ast} \sim\exp[-(d/a)^{c}]$.}
\label{fig4}
\end{figure}

We have simulated the model for other values of the density of fixed links $d$, and we have also observed the above-discussed transition (see Fig. \ref{fig3}). Alternatively, we take into account the transition values $\lambda^{\ast}$ as functions of $d$ for fixed values of $q$ and we have noted that the data obtained from the simulations are well described by a stretched exponential function,
\begin{equation}
\lambda^{\ast}(d) \sim \exp[-(d/a)^{c}],
\label{stretched}
\end{equation}
for different values of the parameters $a$ and $c$ (see Fig.~\ref{fig4}). In particular, for small values of $q$ like $q=0.3$, the dependency of $\lambda^{\ast}$ with $d$ is almost purely exponential, since we have obtained an exponent $c=0.91 \pm 0.08$. The values of $a$ and $c$ are shown in Tab.~1. We have no first principles justification for this dependence yet, but the fairness of such approximation will be justified later on.

\begin{table}
\begin{center}
\begin{tabular}{llll}
\hline
$q$ & $a$ & $c$ & $\chi ^2$ \\ \hline\hline
$0.3$ & $0.62\pm 0.04$ & $0.91\pm 0.08$ & $1.13 \times 10^{-4}$\\
$0.5$ & $0.34\pm 0.01$ & $0.93\pm 0.05$ & $2.46 \times 10^{-4}$\\
$0.8$ & $0.11\pm 0.01$ & $0.73\pm 0.09$ & $6.28 \times 10^{-4}$\\ \hline
\end{tabular}
\end{center}
\caption{Values of the parameters $a$ and $c$ used in Fig.~\ref{fig4}.}
\end{table}

%%%%%%%%%%%%%%%%%%%%%%%%%%%%%%%%%%%%%%%%%%%%%%%%%%%%%%%%%%%%%%%%%%%%%%%%%%
%%%%%%%%%%%%%%%%%%%%%%%%%%%%%%%%%%%%%%%%%%%%%%%%%%%%%%%%%%%%%%%%%%%%%%%%%%%

\begin{figure}[t]
\begin{center}
\vspace{1.0cm}
\includegraphics[width=0.5\textwidth,angle=0]{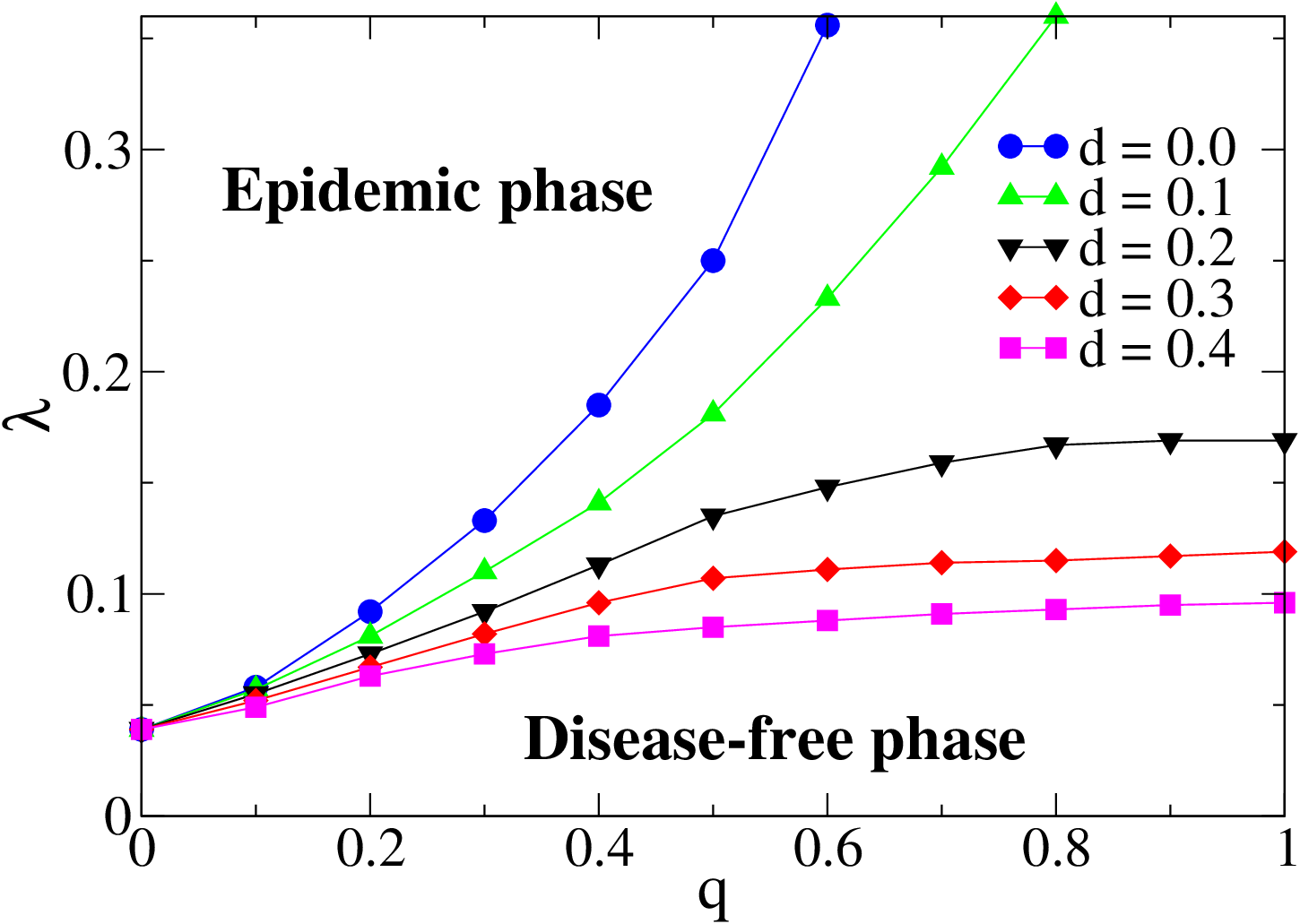}
\end{center}
\caption{Phase diagram of the model, separating the disease-free and the epidemic phases, for typical values of the density of fixed links $d$. The symbols represent the threshold values $\lambda^{\ast}$ obtained from the numerical simulations and the lines are plotted as a guide to the eye.}
\label{fig5}
\end{figure}

Allowing for the transition values $\lambda^{\ast}$ obtained from the simulations, we exhibit the phase diagram of the model in the plane $\lambda$ versus $q$ for typical values of $d$ in Fig.~\ref{fig5}. From this plot, it is possible to understand that a disease-free regime is achievable even for high infection rates and low compliances with the medical recommendations of home isolation, but only when the individuals are near to living alone or the ratio between the members of her household by the total number of relationships is small. As the fraction of fixed links (relationships) $d$ associated with the household soars, the size of the epidemic phase increases as well. This is easily comprehended: since the number of fixed links augments, the quenched part of the network becomes more dominant and so the number of ``channels'' through which the epidemics can disseminate. Correspondingly, we can end up in a situation for which increasing the consciousness $q$ of the infected individual has little or none effect. In this respect, looking at the curve with $d=0.3$, we verify that the threshold infection rate $\lambda^{\ast}$ barely varies with the rate $q$, for $q >\sim 0.6$.

On the other hand, in the interpretation of this diagram, we can also resort to the poll results published by Eastwood and co-workers \cite{eastwood}. Looking to the data from Australia, during the H1N1 outbreak the lowest compliance rate with a public health request was of $62.6\%$ for the avoidance of public gatherings. The rate of compliance decreases further to as much as $50\% $ when household quarantine is considered~\cite{kavanagh}. As previously mentioned, in the same work of Ref.~\cite{kavanagh}, it was given statistical significance to the fact that literacy skills influence the individual performance regarding proper compliance with a self-isolation recommendation as well as other public health interventions. Bearing in mind that Australia is one of the few countries with a Human Development Index (HDI) higher than $0.900$, which in total account for less $10\%$ of the World's population (World HDI = $0.751$), and that it is expected that the lower the HDI, the lower the compliance rates, assuming reasonable values for the parameters we evidence that it is extremely unlikely that a Q\&I policy under such conditions will induce a disease-free state. As a matter of fact, for infection rates of approximately $20\%$, as those estimated for the influenza, and $d=0.2$ (8 acquaintances per 10 relationships and the household composed of three people), we get an epidemic phase if even the probability of compliance is 1, which is basically an utopian value. As regards the value of $d$, we must allow for the fact that at present, people tend to actually (in a vis-\`{a}-vis sense) interact much less than in the past, particularly in countries where internet has had strong implantation. In other words, although the world is apparently more connected, activities like home-working and on-line shopping among others have boosted their relevancy and with that decreasing the number of acquaintances (and casual contacts).

%%%%%%%%%%%%%%%%%%%%%%%%%%%%%%%%%%%%%%%%%%%%%%%%%%%%%%%%%%%%%%%%%%%%%%%%%%
%%%%%%%%%%%%%%%%%%%%%%%%%%%%%%%%%%%%%%%%%%%%%%%%%%%%%%%%%%%%%%%%%%%%%%%%%%%

%##################################################################################
%{\color{blue}

\section{Analytical considerations}

Following the rules of our model, we can do some analytical considerations
regarding our problem, which corresponds to a network of $N$ individuals and
$L$ dual flux links that are quenched for the household and allow rewiring
for the other cases. At initial time, $t_{0}$, the average degree of
connectivity is $K_{0}=2\,L\,/\,N \,=\, \langle k\rangle$ and can be split into%
\begin{equation}
\begin{array}{cl}
K_{0} & =K^{h}+K^{a} \\
&  \\
& =d\,K_{0}+\left( 1-d\right) \,K_{0}.%
\end{array}
\label{link1}
\end{equation}%
Assuming all individuals (nodes) equal, we can write a
master equation reflecting the evolution of the number of infected people
which reads,%
\begin{equation}
\begin{array}{cl}
I_{t+1} & =I_{t}+\lambda \,S_{t}\,\,I_{t}-\alpha \,I_{t} \\
&  \\
& =I_{t}+\lambda \left( K_{t}-I_{t}\right) \,I_{t}-\alpha \,I_{t},%
\end{array}
\label{mf1}
\end{equation}%
which in the continuum limit yields,%
\begin{equation}
\frac{dI}{dt}=\lambda \left( K_{t}-I_{t}\right) \,I_{t}-\alpha \,I_{t}
\label{mf2}
\end{equation}
where,
\begin{equation}
S_{t}+I_{t}=K_{t}.
\end{equation}

When the individual preserves all of its links, $K_{t}=K_{0}$ (for all $t$),
the stability limit of the solution to Eq. (\ref{mf1}) yielding $I_{t\rightarrow \infty }=0$
is given by the infection rate $\lambda =\alpha /K_{0}$~as given in Ref.~\cite{diekmann}. In our case, we must focus on the dynamics of the connections of infected people as well. We know that at each time step, an infected person is given a
medical advice to self-isolate, which implies in a cut of her acquaintances
relations. In view of the fact that diseased individuals have a probability $q$ of disrupting these
contacts, the evolution of $K^{a}$ of an
infected person is on average given by,%
\begin{equation}
\begin{array}{cl}
K_{t+1}^{a} & =K_{t}^{a}-q\,K_{t}^{a} \\
&  \\
\frac{dK^{a}}{dt} & =-q\,K^{a},%
\end{array}
\label{link2}
\end{equation}%
the solution thereto is,%
\begin{equation}
K_{t}^{a}=\left( 1-d\right) \,K_{0}\,\exp \left[ -q\,t\right] ,
\end{equation}%
or,%
\begin{equation}
K_{t}=K_{0}\left[ d\,+\left( 1-d\right) \,\,\exp \left[ -q\,t\right] \right]
.  \label{linkevo}
\end{equation}

Now, the determination of the critical threshold from Eqs.~(\ref{mf2})~and~(\ref{link2})
can be tremendously simplified using the argument~\cite{gross} that
for the epidemic phase to persist in time, one must guarantee the infection
of other people during her illness span that lasts a time of the order of $%
\alpha ^{-1}$. In other words, the reproductive rate of infected people, $R$, must be
greater than one. At time step $t$, we can determine the mean-field reproductive rate,
\begin{equation}
R_{t}=\,K_{0}\frac{\lambda }{\alpha }\left[ d\,+\left( 1-d\right) \,\,\exp %
\left[ -q\,t\right] \right] .
\end{equation}%
Averaging $R_{t}$ during the diseased cycle,%
\begin{equation}
\bar{R}=\alpha \int_{0}^{\alpha ^{-1}}R_{t}\,dt,
\end{equation}%
and solving $\bar{R}=1$ in order to $\lambda $, we get,%
\begin{equation}
\lambda _{c}=\frac{q}{\,K_{0}\left[ 1-d\left( 1-\frac{q}{\alpha }\right)
-\exp \left[ -\frac{q}{\alpha }\right] \left( 1-d\right) \right] }.
\label{critical}
\end{equation}%
the limits of which, namely, $d\rightarrow 1$ (quenched network) and $%
d\rightarrow 0$ (complete reconnection with probability $q$) are verified~\cite{anderson_may,diekmann,gross}.

%%%%%%%%%%%%%%%%%%%%%%%%%%%%%%%%%%%%%%%%%%%%%%%%%%%%%%%%%%%%%%%%%%%%%%%%%%
\begin{figure}[t]
\begin{center}
\vspace{1.0cm}
\includegraphics[width=0.45\textwidth,angle=0]{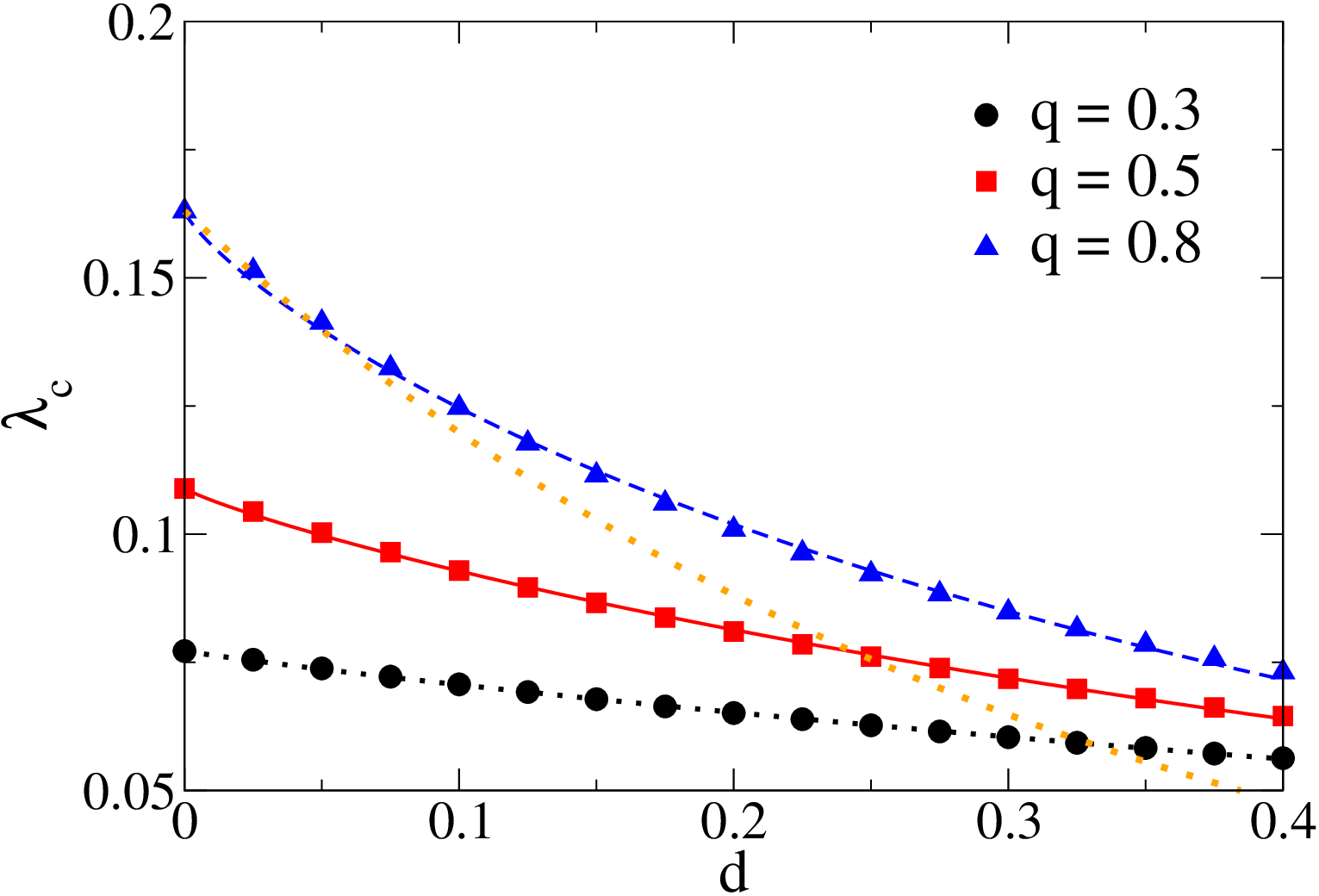}
\includegraphics[width=0.45\textwidth,angle=0]{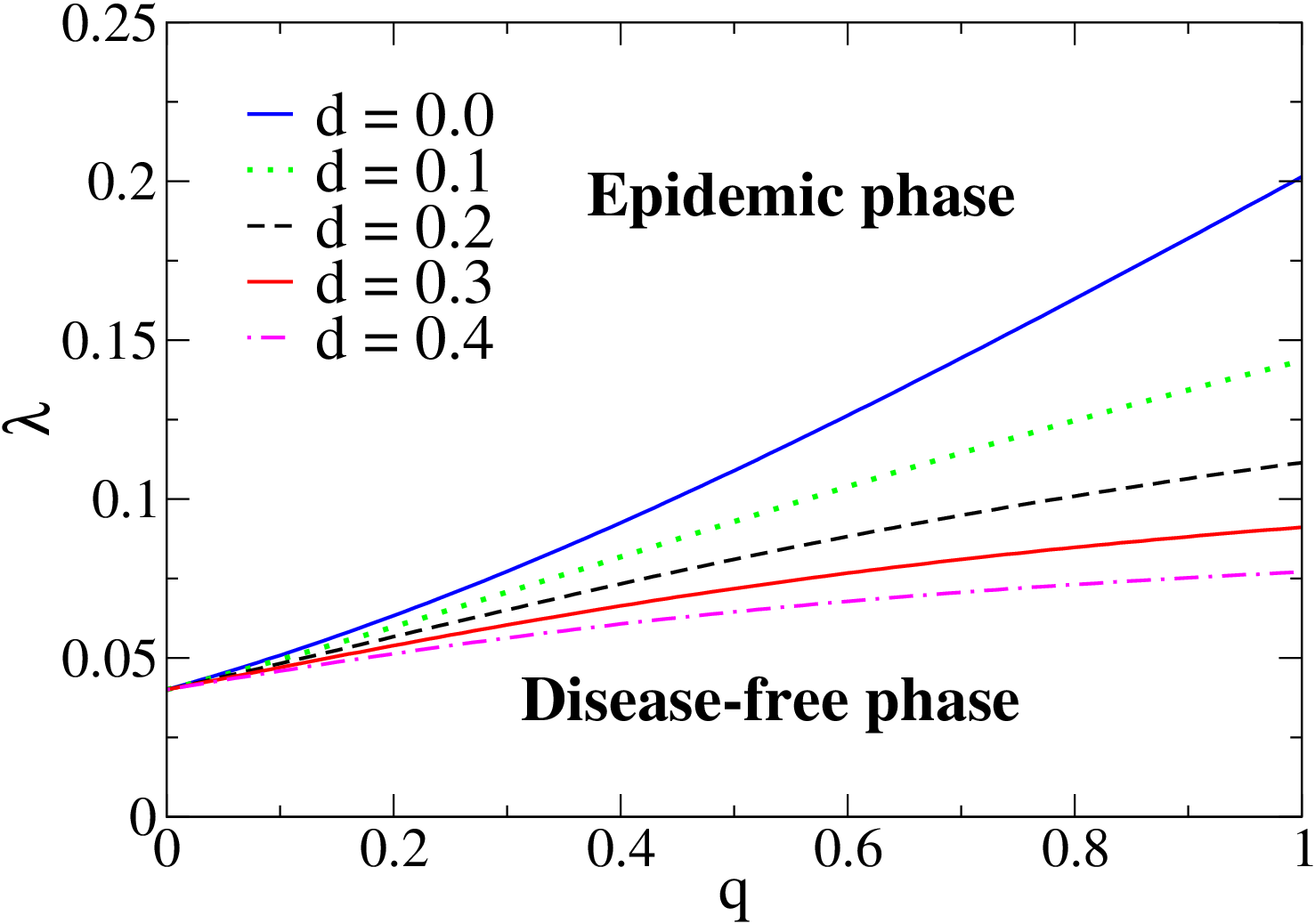}
\end{center}
\caption{
%\color{blue}{
Left: Critical infection probability of\ general solution $\lambda _{c}$ vs
fraction of quenched links $d$ representing the household (symbols) and the numerical adjustment considering Eq.~(\ref{stretched}). Right: Phase diagram of the model considering 
the general solution $\lambda_{c}$ vs the probability $q$ 
of compliance with a self-isolation recommendation. Both curves are obtained
from Eq. (\ref{critical}), using the same parameters we introduced in the
simulations, namely $\alpha =0.2$ and $K_{0}=5$. The colour scheme is the same
as Figs.~\ref{fig4}~and~\ref{fig5} for easier comparison.
}
%}
\label{fig6}
\end{figure}
%%%%%%%%%%%%%%%%%%%%%%%%%%%%%%%%%%%%%%%%%%%%%%%%%%%%%%%%%%%%%%%%%%%%%%%%%%%

In Fig.~\ref{fig6}, we depict $\lambda _{c}$ as a function of $d$ and $q$,
respectively. Comparing both panels with Figs.~\ref{fig4}~and~\ref{fig5}, we verify that only the
\emph{qualitative} behaviour of $\lambda _{c}$ concurs with that of $\lambda ^{*}$.
Specifically, the mean-field solution \emph{underestimates} the critical values of the
infection probability. To understand this discrepancy we avail ourselves of
recently published work~\cite{brits} where it is ascertained that mean-field
like approaches are more accurate for networks with high mean degree or high
mean first neighbour degree. As can be perceived from Eq.~(\ref{linkevo}) and using
the same parameters values of the simulations, the mean degree significantly fluctuates from susceptible to infected nodes.
Therefore, at the stationary critical state and beyond it, the typical
configuration of the network includes an important fraction of nodes with a
small degree of neighbours, a situation that tallies with conditions of
inaccuracy we have referred to.

Nevertheless, we can provide an argument to our empirical
adjustment of $\lambda ^{*}$ as a function of $d$ from Eq.~(\ref{critical}). Explicitly, for $q>\alpha
$, that equation can be rewritten as,
\begin{equation}
\lambda _{c}=\frac{q}{\,K_{0}\left( 1-\exp \left[ -\frac{q}{\alpha }\right]
\right) }\left[ 1+\frac{\exp \left[ -\frac{q}{\alpha }\right] +\frac{q}{%
\alpha }-1}{1-\exp \left[ -\frac{q}{\alpha }\right] }d\right] ^{-1},
\end{equation}
which in first order yields,
\begin{equation}
\lambda _{c}\sim \exp \left[ -\frac{d}{\mathcal{D}}\right] ,
\label{lambda1order}
\end{equation}
with,
\begin{equation}
\mathcal{D=}\frac{1-\exp \left[ -\frac{q}{\alpha }\right] }{\exp \left[ -%
\frac{q}{\alpha }\right] +\frac{q}{\alpha }-1}.
\end{equation}
A comparison between Eqs.~(\ref{critical})~and~(\ref{lambda1order}) is presented in the left panel of Fig.~\ref{fig6}. 
Therein, for small $d$ the orange dotted line fits the initial points, but then decays faster. The stretching exponent 
$c < 1$ prolongs the curve. We can also confront our proposal Eq.~(\ref{stretched}) with Eq.~(\ref{critical}). 
In Fig.~\ref{fig6} (left side), we have numerically adjusted the points given by Eq.~(\ref{critical}) with  Eq.~(\ref{stretched}). 
As visible, the adjustment matches the curve quite well with hardly perceptible deviations. Because of the qualitative similarity 
between both curves, we can be further tempted to consider an adjustment of the data in Fig.~\ref{fig4} with a mean-field like dependence,
\begin{equation}
\lambda ^{*} = \frac{b}{1+ \,d \,/ \, a ^{\prime }},
\label{fit2}
\end{equation}
the values of which are presented in Tab.~2. Comparing the values of $\chi ^2$ in both tables, we verify that despite the 
good agreement in the mean field curves, the stretched exponential clearly beats Eq.~(\ref{fit2}), especially for large $q$,
which upholds once again our previous choice.

\begin{table}
%{\color{blue}
\begin{center}
\begin{tabular}{llll}
\hline
$q$ & $a ^{\prime }$ & $b$ & $\chi ^2$ \\ \hline\hline
$0.3$ & $0.44\pm 0.02$ & $0.13\pm 0.01$ & $4.5 \times 10^{-4}$\\
$0.5$ & $0.22\pm 0.01$ & $0.25\pm 0.02$ & $4.7 \times 10^{-4}$\\
$0.8$ & $0.05\pm 0.01$ & $0.89\pm 0.04$ & $1.7 \times 10^{-2}$\\ \hline
\end{tabular}
\end{center}
\caption{Values of the parameters $a ^{\prime }$ and $b$ to numerically adjust the points of Fig.~\ref{fig4} with Eq.~(\ref{fit2}).}
%}
\end{table}

%}
%END BRACKET OF COLOR BLUE

%###################################################################################

%###################################################################################
\section{Final Remarks}

\qquad In this work, we have studied a modified version of the SIS epidemic model that takes into account quarantine and isolation practices according to a self-compliance with the medical requests. For each person, their relationships are pigeonholed as acquaintances and members of the same household. For the former, there is a probability that the connection is disrupted, quantifying the degree of consciousness of the individual towards the epidemic disease, whereas for the latter the connections are fixed aiming at representing the possibility of a compulsory estrangement. For a Susceptible-Infected-Susceptible process, our results have shown that quarantines following this structure, which agrees to WHO directives, is extremely unlikely to thwart the propagation of an epidemic infectious disease using standard figures regarding infection rate and compliance with medical (public health) requests. It should be stressed that our results must not be interpreted as arguing the uselessness of self-isolation in a overall point of view. The outcome of our study indicates that self-isolation is likely to be ineffective \emph{per se} and signals the importance of coordinated public health or non-pharmaceutical interventions in order to control the untamed spread of an epidemic.

That being said, we would like to highlight that our results can be further explored to take into account either different versions considering the impact of other public health (non-pharmaceutical interventions) or the joint impact of self-isolation and some other intervention(s). Besides scrutinising different types of epidemics, we can also study the problem of isolation using mixed topologies \cite{moreno} for the social network which depend on the type of relationship between people and institutions. Regarding this point, the problem can be honed by bolstering the acquaintance type connections, \emph{i.e.}, in the rewiring step people are biased to rewire to acquaintances and members of the same household of people with whom they already have a relationship. Another important investigation concerning the dynamics respects the assumption of an (average) isolation time instead of a fixed (unitary) time scale we have taken into consideration.

\section*{Acknowledgments}
We acknowledge J~F~F~Mendes for having provided the computational resources of the Group of Complex Systems and Random Networks of the Aveiro University, Portugal, where part of the simulations were performed. We have benefited from the financial support by the Brazilian funding agencies FAPERJ, CAPES and CNPq (NC) as well as the European Commission through the Marie Curie Actions FP7-PEOPLE-2009-IEF (contract nr 250589) (SMDQ).

\end{document}